# MODELING OF THE GAS-PHASE OXIDATION OF CYCLOHEXANE

*Frédéric BUDA, Barbara HEYBERGER, René FOURNET, Pierre-Alexandre GLAUDE, Valérie WARTH, Frédérique BATTIN-LECLERC**

Département de Chimie-Physique des Réactions, UMR 7630 CNRS, INPL-ENSIC, 1, rue Grandville, BP 20451, 54001 NANCY Cedex - France

ABSTRACT

This paper presents a modelling study of the oxidation of cyclohexane from low to intermediate temperature (650-1050 K) including the negative temperature coefficient (NTC) zone. A detailed kinetic mechanism has been developed using computer-aided generation. This comprehensive low temperature mechanism involves 513 species and 2446 reactions and includes two additions of cyclohexyl radicals to oxygen, as well as, subsequent reactions. The rate constants of the reactions involving the formation of bicyclic species (isomerizations, formation of cyclic ethers) have been evaluated from data of the literature. This mechanism is able to reproduce satisfactorily experimental results obtained in a rapid compression machine, for temperatures ranging from 650 to 900K and in a jet stirred reactor from 750 to 1050 K. Flow rates analyses have been performed at low and intermediate temperatures.

---

* E-Mail : Frederique.Battin-Leclerc@ensic.inpl-nancy.fr, Tel : 33 3 83175125, Fax : 33 3 83378120

# INTRODUCTION

As mentioned by Walker and Morley (1), cycloalkanes have a significant presence in conventional fuels (up to 3 % in gasoline and up to 35 % in Diesel fuel), but relatively little attention has been paid to the chemistry involved during their oxidation. Furthermore, cycloalkanes are key compounds of the formation of aromatic pollutants. Up to now, two detailed mechanisms have been proposed to model the oxidation of cyclohexane. Klai and Baronnet (2,3) have written a low temperature mechanism to reproduce their experimental results obtained in a static reactor at 635 K, at low pressure (6 kPa) and for an equivalence ratio equal to 9, conditions which are rather far from those actually observed in engines. More recently, a team in Orléans (4, 5) has proposed a high temperature mechanism to explain experimental results obtained in jet-stirred reactor for temperatures ranging from 750 to 1200 K, pressures from 1 to 10 atm and equivalence ratios from 0.5 to 1.5. This mechanisms was also validated at higher temperature by modeling the laminar flame speeds of cyclohexanes/air flames (6), but it would not model the oxidation of cyclohexane at lower temperature in the NTC zone.

Granata *et al.* (7) have proposed a globalized mechanism to model the oxidation of cyclohexane, which have been validated using the results of the literature (4-8), including those obtained by Lemaire *et al.* (8) in a rapid compression machine under conditions close to those actually observed in engines and which cover the cool flames and NTC zone: temperatures ranging from 650 to 900K, pressures from 7 to 17 bar, equivalence ratio equal to 1. The purpose of our paper is to reinvestigate the modelling of Granata et al. (7), which was performed with a primary mechanism containing 2 globalized steps to reproduce the high temperature decomposition and 6 reactions to take into account the low temperature behavior, by using a comprehensive primary mechanism generated by an automatic procedure and taking into account all the possible elementary steps for every isomers of peroxy and hydroperoxy radicals. This is also a first step in the determination of specific rate constants for reactions of other cyclic compounds, such as branched cyclic alkanes.

## THE GENERATION OF THE MODEL

According to the literature (1), the elementary steps included in the models of the oxidation of cyclic alkanes are close to those proposed to describe the oxidation of acyclic alkanes. Consequently, it has been possible to obtain the model proposed here for the oxidation of cyclohexane by using an improved version of the software EXGAS, a computer package developed to perform the automatic generation of detailed kinetic models for the gas-phase oxidation and combustion of linear and branched alkanes. As this package and its application to model the oxidation of a wide range of alkanes (9) and several alkenes (1-pentene, 1-hexene (10)) have been already described, the general features of EXGAS are only very briefly summarized in the first part of this chapter.

Nevertheless, the modeling of the oxidation of cyclic alkanes requires to consider new types of generic reactions, and especially to define new correlations for the estimation of the rate constants. Thus the rate parameters for isomerizations and the formations of cyclic ethers are much affected by the presence of a cycle in the reactant. The second part of this chapter details the reactions and the estimations of rate constants, which are specific to the oxidation of cyclic alkanes.

### *General features of EXGAS*

EXGAS provides reaction mechanisms made of three parts :

- A $C_0$-$C_2$ reaction base including all the reactions involving radicals or molecules containing less than three carbon atoms (11) coupled with a reaction base for $C_3$-$C_4$ unsaturated hydrocarbons (12), such as propyne, allene or butadiene, including reactions leading to the formation of benzene,
- A comprehensive primary mechanism, where the only molecular reactants considered are the initial organic compounds and oxygen,

The primary mechanism includes only elementary steps; the reactions, which are considered to model the oxidation of alkanes are the following :

- Unimolecular and bimolecular initiation steps,
- Decomposition and oxidation of alkyl radicals (to form the conjugated alkene plus $HO_2$•),
- Addition to oxygen of alkyl and hydroperoxyalkyl radicals,

- ♦ Isomerizations of alkyl and peroxy radicals,
- ♦ Decompositions of hydroperoxyalkyl and di-hydroperoxyalkyl radicals to form cyclic ethers, alkenes, aldehydes, ketones and oxohydroperoxyalkanes,
- ♦ Metatheses involving the H-abstraction reactions from the initial reactants by a radical,
- ♦ Termination steps.

➢ A lumped secondary mechanism, containing reactions consuming the molecular products of the primary mechanism, which do not react in the reaction bases (13).

Thermochemical data for molecules or radicals are automatically calculated and stored as 14 polynomial coefficients, according to the CHEMKIN II formalism (14). These data are calculated using software THERGAS (15), based on the group and bond additivity methods proposed by Benson (16). Thermochemical data for cyclic molecules and radicals included in the primary mechanism of the oxidation of cyclohexane are shown in Table 1.

TABLE 1

The kinetic data used in the reaction bases were taken from the literature; values recommended by Baulch *et al.* (18) and Tsang *et al.* (19) are mainly used in the $C_0$-$C_2$ reaction base; the pressure-dependent rate constants follow the formalism proposed by Troe (20) and efficiency coefficients have been included. The kinetic data of the reactions included in the primary or secondary mechanisms are either calculated using thermochemical kinetics methods or estimated using correlations (9, 13).

*Reactions and rates constant specific to cyclic alkanes*

On a computing point of view, the treatment of cyclic reactants has induced important modifications, as the tree-like internal notation used by the programme had to be improved to take into account the presence of a cycle, with a new algorithm of canonicity to avoid redundant products or generated reactions (21).

On a chemical point of view, every generic reaction involved in the primary mechanism, as well as the associated rate parameters, have been analyzed to take into account the differences induced by the presence of the cycle. Table 2 shows the primary mechanism related to the species containing a 6-members ring. The remaining of the primary mechanism, which is the biggest part, is related to linear species obtained after opening of the ring and has been generated using previously described rules (9, 10).

TABLE 2

Unimomecular initiation steps

Taking into account the range of temperature studied (below 1050 K), the reactions leading to the formation of biradicals have not been considered. The only unimolecular initiation is then the formation of cyclohexyl radicals by breaking of a C-H bond, for which we have considered the reverse reaction (reaction 1 in table 3) with a rate constant ($k = 1.0 \times 10^{14} s^{-1}$) according to Allara *et al.* (22). The kinetic role is played by the reverse reactions as a termination step. Simulations showed that the unimolecular initiation involving the breaking of a C-C bond and the formation of a diradical and leading mainly to the formation of 1-hexene, which is of importance to model results above 1200 K (23), can here be neglected.

H-abstraction reactions from the reactant (bimolecular initiations and metatheses)

As the bond dissociation energy of the C-H bond in cyclohexane (99.5 kcal/mol) is not very different than that of a secondary C-H bond in n-butane (99.3 kcal/mol), the rate constants for H-abstraction have been deduced from those of the abstraction of secondary H-atoms in acyclic alkanes (9). The same rate constant has been used for the bimolecular initiation (reaction 2); but, to obtain a good agreement for the simulations in a jet-stirred reactor, the A-factors for the metatheses with H atoms, OH and HO2 radicals from cyclohexane to give cyclohexyl radicals (reactions 46-48) have been considered five times higher than in the case of alkanes (9). In the case of the metatheses with H atoms and OH radicals, the rate constants used here are in agreement within a factor 1.8 and 5, respectively, with the values measured by Gulati *et al.* (24) at 753 K.

Reactions with oxygen molecules

Considering that the presence of the cycle do not strongly modify the reactivity of alkyl radical towards oxygen, the rate parameters used have been deduced from the case of acyclic alkanes (9), with a A factor 3 times lower for the addition to oxygen of cyclohexyl and hydroperoxycyclohexyl radicals (reactions 3-7) ($k = 3.5x10^{18}\ T^{-2.5}\ cm^3.mol^{-1}.s^{-1}$, when the radical center neighbors a carbon atom carrying an oxygen atom, $k = 6x10^{18}\ T^{-2.5}\ cm^3.mol^{-1}.s^{-1}$, for the other cases). We have used the same rate constant for the oxidation of cyclohexyl radicals to give cyclohexene (reaction 45) as for alkyl radicals ($k = 3.9x10^{12}$ exp (-5/RT) $cm^3.mol^{-1}.s^{-1}$). Throughout this paper R is the gas constant expressed in kcal $mol^{-1}\ K^{-1}$.

Isomerisations of cyclic peroxy and peroxy hydroperoxyalkyl radicals

Isomerizations (reactions 8-30) involve an internal transfer of a hydrogen atom to form a new hydroperoxy group and occur through a bicyclic transition state, which involves the formation of an additional ring including from 4 to 7 members. The related rate parameters (k = exp(-Ea/RT) are presented in Table 3 for peroxy cyclohexyl radicals. A-factors are mainly based on the changes in the number of losses of internal rotations as the reactant moves to the transition state (here one rotation is lost in every cases) and on the reaction path degeneracy (here the number of transferable atoms of hydrogen) (13). As presented previously for alkylperoxy radicals (9, 13), the activation energy is equal to the sum of the strain energy of the cyclic transition state and the activation energy for H-abstraction from the substrate by analogous radicals. The activation energy for H-abstraction is that of the abstractions of secondary H-atoms in acyclic alkanes (i.e. 17 kcal/mol) (9). The strain energy of a compound can be calculated as the difference between the experimental enthalpy (ΔH°(exp)) and the sum of the contributions of groups as proposed by Benson (ΔH°(groups of Benson)).

TABLE 3

In the case of the isomerisation involving a transition state with a 6 members ring (see table 3), the strain energy for the ring including two oxygen atoms (15.8 kcal/mol) is deduced from that of a model compound with no oxygen atom by adding the same difference as for acyclic alkanes (i.e. 8 kcal/mol (9)). The considered model bicyclic compound is then bicyclo[3,3,1]nonane, the strain energy of which can be estimated as 7.8 kcal/mol, since ΔH°(exp) is equal to -30.5 kcal/mol (25) and ΔH°(groups of Benson) to -38.3 kcal/mol (16). In the case of the isomerisation involving a transition state with a 4 members ring, the influence of the second ring is neglected; the strain energy is that of the monocyclic transition state (9). The activation energy for H-abstraction is that of the abstractions of tertiary H-atoms neighboring an oxygen atom (i.e. 12 kcal/mol) (9).

In the case of the isomerisations involving a transition state with a 5 members ring, the activation energy is deduced from A, the rate constant ($k_6$) for the isomerization involving a transition state with a 6-members ring and the ratio between $k_6$ and the rate constant ($k_5$) of the isomerization involving a transition state with a 5-members ring ($k_5/k_6 = 0.19$) measured by Walker et al. at 753 K (1, 24). The A-factor is divided by 200, to take into account the ratio of molecules which are in the "boat" form, instead of in the "chair" one (i.e. 0.5 % (1)).

In the case of the isomerisations involving a transition state with a 7 members ring, the activation energy is deduced from A, ($k_5$) and the ratio between $k_5$ and the rate constant ($k_7$) of the isomerization involving a transition state with a 7-members ring ($k_5/k_7 = 0.60$) measured by Walker et al. at 753 K (1, 24).

It is worth noting that, because of the formation of bicyclic transition states, the activation energies presented in Table 3 are higher than for acyclic compounds. For instance, for a transition state with 6 members, the activation is 8 kcal/mol higher for cyclohexane than for acyclic alkanes and the rate constant at 750 K is then around 200 times lower for the cyclic alkane.

Table 3 also shows that at 753 K, there is a very good agreement between the rate constants calculated using our rate parameters and those proposed by Handford-Styring and Walker between 673 and 773 K (26).

As proposed by Glaude *et al.* (27), we consider for peroxy hydroperoxycyclohexyl radicals, a direct isomerization-decomposition to give ketohydroperoxides and hydroxyl radicals. For reducing the number of reactants in the secondary mechanism (13), the different isomers of a primary molecular product are lumped into one unique species: this is the case for the different isomers of hydroperoxyhexanone obtained by these direct isomerization-decompositions (reactions 12-30).

Decompositions by β-scission

The decompositions by β-scission involving the opening of the ring (reactions 31 and 35-40) have been considered, with the same rate constants as for β-scission involving the breaking of a C-C bond in acyclic alkyl radicals to form a primary radical ($k = 2x10^{13}exp (-28.7/RT) s^{-1}$ for each bond which can be broken (9)). The decompositions by beta-scission involving the breaking of a C-H bond from cyclohexyl leads to the formation of cyclohexene. We have considered the reverse reaction, the corresponding addition of H atom to the double bond (reaction 32), which has a rate constant better defined ($k = 2.6x10^{13} exp(-1.5/RT) cm^3.mol^{-1}.s^{-1}$ (28)). The decompositions by β-scission of hydroperoxy cyclohexyl radicals to give cyclohexanone and OH radicals (reaction 33) or cyclohexene and $HO_2$ radicals (reaction 34) have been considered with the same rate constant as for acyclic compounds (9). The rate constant of the decomposition by β-scission of the cyclic alcoxy radicals to give 1-hexanal-6-yl (reaction 41) has been taken equal to that the decomposition of linear alcoxy radicals (29). The A-factor is assumed to be the same as for a β-scission of alkyl radicals; the activation is estimated using thermochemistry from the activation energy of the reverse reaction, which is about 10 kcal/mol according to Benson (16). The reactions deriving from 1-hexen-6-yl and of 1-hexanal-6-yl radicals have been generated by EXGAS using the same rules as the ones which were validated by modeling the oxidation of linear alkanes (9) and alkenes (10), including the additions to oxygen, which were neglected by Granata *et al.* (7).

Decompositions to give cyclic ethers

Hydroperoxycycloalkyl radicals can decompose to give bicyclic ethers with an additional ring including from 3 to 5 members (reactions 42-44). The ring strengths of the respective transition states are then very different from that involved in the reaction of linear or branched peroxy radicals. The activation energies have been evaluated from theoretical density function theory calculations performed with Gaussian03 software (30). The cyclic peroxy radicals and the transition states have been calculated at the B3LYP/cbsb7 level. Intrinsic Reaction Coordinate calculations have been performed to insure that the transition states connect correctly the reactants and the products. A-factors have been mainly based on the changes in the number of losses of internal rotations induced by the formation of the cyclic ether (here one rotation is lost in every cases), but have been divided by 3 compared to the formula given by Warth et al. (13). According to Walker et al. (1, 24), the bicyclic ether with an additional ring involving 4 members is very unstable and decompose very quickly to 1-hexenal. The reaction (43) transforms then directly the peroxy radical to 1-hexenal. The related rate parameters (k = Axexp(-Ea/RT) are presented in Table 4.

TABLE 4

Disproportionnation of hydroperoxy alkyl radicals

The disproportionnation between hydroperoxy cyclohexyl and $\bullet HO_2$ radicals (reaction 57) has been considered with the same rate constants as for acyclic compounds ($k = 2x10^{11}exp(1.3/RT)$ $cm^3.mol^{-1}.s^{-1}$ (9)). The low reactivity of cyclic hydroperoxy alkyl radicals towards isomerizations has required to take also into account the disproportionnations of hydroperoxy cyclohexyl radicals with themselves (reactions 58, 59) and with small peroxy radicals such as $CH_3OO$, $C_2H_5OO$ and $iC_3H_7OO$ radicals (reactions 60-65). The following reactions have been considered, with rate constants derived from those proposed for the disproportionnations of $CH_3OO$ radicals with themselves by Walker *et al.* (1) :

- Disproportionation of two hydroperoxy alkyl radicals to give radicals

  (e.g. 2 cyclic-$C_6H_{11}OO\bullet \Rightarrow$ 2 cyclic-$C_6H_{11}O\bullet + O_2$), with $k = 6.3x10^{10}exp(0.725/RT)$ $cm^3.mol^{-1}.s^{-1}$,

- Disproportionation of two hydroperoxy alkyl radicals to give molecules

  (e.g. 2 cyclic-$C_6H_{11}OO\bullet \Rightarrow$ cyclic-$C_6H_{11}OH$ + cyclic-$C_6H_{11}O + O_2$), with $k = 1.4x10^{10}exp\ (0.725/RT)$ $cm^3.mol^{-1}.s^{-1}$.

These types of reaction, which explain the formation of cyclohexanol observed by Klai and Baronnet (2), were not considered by Granata *et al.* (7)

Secondary reactions of cyclohexene

The recently published mechanism for the oxidation of cyclohexene at high temperature (31) has been added in order to better consider the secondary reactions of this molecule. This mechanism does not consider addition of radicals to oxygen molecules, but involves molecular eliminations, unimolecular and bimolecular initiations, addition of small radicals to the double bond, radical decompositions by the break of a C–C or a C–H bond, oxidation involving the H-abstraction by $O_2$ from a radical yielding the conjugated alkene, H-abstraction by O, H, OH, $HO_2$ and $CH_3$ free radicals from cyclohexene producing allylic and alkenyl cyclic radicals and combinations of small radicals H, OH, $CH_3$ and $HO_2$ with cyclic radicals, which are yielded from H-abstraction from the reactant. This mechanism includes also a secondary mechanism for the reaction of cyclohexadiene and is connected to our recently published mechanism for the oxidation of benzene (32).

## COMPARISON BETWEEN COMPUTED AND EXPERIMENTAL RESULTS

Simulations have been performed using the software SENKIN of CHEMKIN II (14). In all the figures presented below, the points refer to experimental observations and the curves come from simulations. In order to keep this paper reasonably short, the complete mechanism, which involves 523 species and 2446 reactions, is not presented here, but it is available on request.

Lemaire *et al.* (8) have studied the oxidation and the auto-ignition of cyclohexane in a rapid compression machine under the following conditions after compression : temperatures between 600 and 900 K and pressures from 7 to 14 bar. The total autoignition delay times were measured between the end of compression and the sharp and important rise of pressure due to the final ignition, whereas the cool flame delay times were measured between the end of compression and the maximum intensity of the associated light emission. Concentration-time profiles of the $C_6$ intermediate products of oxidation were also obtained the time before auto-ignition. Figure 1 displays a comparison between the computed and the experimental cool-flame and ignition delay times vs. temperature for an equivalence ratio equal to 1 and for an initial pressure before compression equal to 0.467 bar (fig. 1a) and to 0.730 bar (fig. 1b).

Simulated delay times were deduced from the rises observed in the computed profiles of pressure. These figures show that our model is able to reproduce satisfactorily the cool flame delay times and the ignition delay time at temperatures lower than 700 K, which implies that our primary mechanism is meanly correct. Nevertheless, simulations show a too marked negative temperature coefficient (NTC), which may be due to problems in our secondary mechanism, e.g. the secondary mechanism for cyclohexene does not consider addition of radicals to oxygen molecules as it had to be written to model data at high temperature (31).

FIGURE 1

Figure 2 presents a comparison between experimental results and simulations for the conversion of cyclohexane and the productions of intermediate $C_6$ products analyzed by Lemaire *et al.* (8). As it is often the case in rapid compression machine experiments, the conversion of cyclohexane during the cool flame is strongly overestimated by the model. That is due to the gradient of temperature observed in a rapid compression machine: our simulation refers to the consumption of reactant in the area of maximum temperature, while the experimental value is averaged considering all the temperature zones. Despite this disagreement in the global reactivity, it is interesting to compared the experimental and simulated products distribution: figure 2 also presents the experimental and simulated selectivity of different products which have been analysed. The selectivity, $S_i(t)$, in % molar of a species i, at a residence time t, is given by the following equation :

$$S_i(t) = 100 \times n_i(t)/(n_r(t_0) - n_r(t))$$

with : $n_i(t)$: Number of mole of the species i at the residence time t,

$n_r(t_0)$: Numbers of mole of reactant at the initial time $t_0$,

$n_r(t)$: Numbers of mole of reactant at the residence time t.

Figure 2 shows that the relative selectivities of the main primary products, i.e. cyclohexene, cyclohexadiene, benzene, 1,2-epoxycyclohexane, 1,4-epoxycyclohexane and hexenal, are well computed.

FIGURE 2

Voisin *et al.* (4) have presented experimental results obtained for the oxidation of cyclohexane in a jet-stirred reactor between 750 and 1050 K, at 10 bar, at a residence time of 0.5s and at three equivalence ratios ($\phi$), 0.5, 1, and 1.5. Figures 3 to 5 present a comparison between the experimental and simulated profiles of species for the three equivalence ratios respectively. The model reproduces well the consumption of cyclohexane and oxygen and the formation of the major $C_6$ primary products, cyclohexene and hexenal, except at $\phi$=0.5, for which the formation of hexenal is overestimated by a factor around 2. The production of cyclohexadiene and benzene, which are secondary products deriving from cyclohexene, is also satisfactorily modelled, as well as that of ethylene, which is the main primary products deriving from the opening of the cyclohexyl ring. The computed formation of 1,3-butadiene, which is also obtained from the opening of the cyclohexyl ring, is overestimated, especially at $\phi$=1.5. The agreement concerning light aldehydes (formaldehyde, acetaldehyde, acroleine) and carbon oxides is globally acceptable, as it is in most cases better than a factor 1.5.

FIGURES 3 TO 5

## DISCUSSION

Figures 6 to 8 present flow rates analyses performed under conditions related to the rapid compression machine (at 650 and 750 K) and the jet-stirred reactor (at 900K) experiments, respectively. Figure 9 displays sensitivity analyses for both types of reactor. Under every conditions, cyclohexane is mainly consumed by H-abstraction by small radicals (mainly OH radicals under the conditions of figures 6 and 7) to give cyclohexyl radicals, the fate of which depends strongly on temperature.

FIGURES 6, 7 AND 8

<u>Reactions of cyclohexyl radicals at low temperature (figures 6 and 7)</u>

At 650 and 750 K in a rapid compression machine, cyclohexyl radicals react mainly with oxygen either by oxidation (19 % of the consumption at 650 K, 36 % at 750 K) to form cyclohexene and $HO_2$ radicals or by addition (80 % at 650 K, 60 % at 750 K) to give cyclohexyl peroxy radicals, the fate of which depends also much on temperature. The sensitivity analysis of fig. 9a shows that the rate constant of the additions to oxygen is the most sensitive parameter, especially at the lowest temperatures. An important overprediction of the reactivity has been obtained when attempting to model the results of Klaï and Baronnet (2,3) obtained in a static reactor at 635 K and 6 kPa, which seems to be due to too fast additions to oxygen in these conditions. The rate constants that we propose here for the addition of $O_2$ to form the collisionally stabilized peroxy radical and for the competing formation of alkene and $HO_2$ radicals are satisfactory at the average pressure of this study (around 10 bar), but are certainly pressure dependent and should be used with care at much lower pressures. At 650 K, 80% of cyclohexyl peroxy radicals react by disproportionations: 13 % by reaction with $HO_2$ radicals to give cyclohexane hydroperoxide, 12 % to produce cyclohexanone, cyclohexanol and oxygen molecules and 55 % to give oxygen molecules and cyclohexyl alcoxy radicals, which decompose by opening of the ring to give 6-yl-1-hexanal radicals leading to ketohydroperoxides by successive additions to oxygen and isomerisations. The hydroperoxides obtained can easily decompose by breaking of the O-OH bond inducing a multiplication of the number of radicals and explaining the high reactivity observed at low temperature. This importance of disproportionations is not observed for acyclic hydrocarbons and is induced by a lower rate of isomerizations of peroxy radicals due to strained bicyclic transition states. This low reactivity peroxy radicals by isomerizations induces also an important promoting effect on cool flame delay times of the H-abstractions by cyclohexylperoxy radicals from cyclohexane (1% of the flux of consumption of the these peroxy radicals), as shown in figure 1a by a simulation in which the rate constant of this reaction has been multiplied by 5.

At 750 K, 76% of cyclohexyl peroxy radicals react by isomerisation: 3% by an isomerisation involving a 4-members ring and leading to cyclohexanone, 10% by an isomerisation involving a 5-members ring and producing 1,2-epoxycyclohexane, 50% by an isomerisation involving a 6-members ring and producing hexenal and 13% by an isomerisation involving a 7-members ring and producing 1,4-epoxycyclohexane. Only a small fraction of the hydroperoxy cyclohexyl radicals obtained by isomerisation add to oxygen and involves the formation of ketohydroperoxides. The formation of hydroperoxides is then more reduced than at lower temperature inducing a decrease of the reactivity of the mixture and the appearance of the negative temperature coefficient zone experimentally obtained. That explains also the fact that isomerisations have an inhibiting effect above 700 K, as shown in figure 9a, while in the case of alkanes, they have a pronounced promoting effect (9). The sensitivity analysis for the other reactions displays the same results as in the case of alkanes (9). Cyclohexene is consumed by secondary reactions to give cyclohexadiene and then benzene.

Reactions of cyclohexyl radicals at high temperature (figure 8)

In a jet-stirred reactor at 900K, the reactions of cyclohexyl radicals with oxygen are quite less important (the oxidation accounts for 7% of their consumption and the addition to 6%), while the decompositions either by breaking of a C-H bond (31 %) or by opening of the cycle to give 1-hexene-6-yl radicals (56 %) are the major channels.

The two major primary $C_6$ products are then cyclohexene, which is formed from cyclohexyl radicals by oxidation and decomposition by breaking of a C-H bond, and hexenal, which is obtained from the addition of cyclohexyl radicals to oxygen. Cyclohexene is mainly consumed by H-abstractions to give a resonance stabilized radical which react by oxidation with oxygen to give cyclohexadiene and $HO_2$ radicals or by combination with $HO_2$ radicals to produce ethylene, carbon monoxide and OH and allyl radicals. Cyclohexadiene is consumed by H-abstractions to produce resonance stabilized radicals which react with oxygen to give benzene and $HO_2$ radicals. Hexenal is consumed by H-abstraction to form $C_5H_9CO$ radicals, which either decompose to give carbon monoxide, ethylene and allyl radicals or, to a lesser extend, add to oxygen and finally lead to the formation of pentadiene, carbon dioxide and OH radicals. Allyl radicals mainly react with $HO_2$ radicals to give acroleïne and OH radicals.

The 1-hexene-6-yl radicals obtained by the opening of the cycle rapidly isomerize to form resonance stabilized 1-hexene-3-yl radicals which can either react by oxidation to produce 1,3-hexadiene and $HO_2$ radicals or by decomposition to give 1,3-butadiene and ethyl radicals, which are the main source of ethylene. 1,3-butadiene is mainly consumed by addition of OH radicals to give formaldehyde and allyl radicals or acetaldehyde and vinyl radicals. The deteriorated prediction of the concentrations of 1,3-butadiene and hexenal by the model is probably due to uncertainties in the mechanism of their consumption. As shown in figure 9b, the rate constants of the metatheses are the most sensitive parameters. It is worth noting that the additions to oxygen have still a significant promoting effect at 850 K. The oxidations leading to cyclohexene and $HO_2$ radicals have a promoting effect in these conditions, because easy initiations can occur from this cyclene to produce resonance stabilized radicals.

## CONCLUSION

This paper presents a new detailed kinetic mechanism for the oxidation of cyclohexane at low and intermediate temperature. This mechanism has been obtained using an improved version of software EXGAS in which the generic reactions and rate parameters specific to cyclic alkanes have been implemented. Special care has been taken to evaluate the rate constants for the isomerizations and the formations of cyclic ethers involving the formation of bicyclic species. A comprehensive mechanism of the oxidation of the radicals obtained by opening of the ring has also been included.

This model has allowed us to reproduce satisfactorily cool flames times and ignition delay times below 700 K and selectivities of primary products measured in a rapid compression machine between 650 and 900K, as well as mole fraction profiles obtained in a jet stirred reactor from 750 to 1050 K. That shows the correctness of our primary mechanism, but future work are needed to improve the secondary mechanism of cyclohexene at low temperature in order to better reproduce ignition delay times in the negative temperature coefficient area.

Flow rate and sensitivity analyses show that the two additions of cyclohexyl radicals to oxygen are important to explain the autoignition at low temperature, whereas at intermediate temperature, the decomposition of cyclohexyl radicals by beta-scission should also be considered. At 650 K, an unusual importance of the disproportionations of peroxy radicals is observed, which is induced by a lower rate of isomerizations of peroxy radicals due to bicyclic transition states.

It would also be of interest to develop such a model for cyclopentane and for substituted cyclic alkanes (methyl or propyl cyclohexane), but experimental data to be used for validation are still scarce.

ACKNOWLEDGMENTS

This work has been supported by the program ECODEV of the CNRS, by TOTAL and PSA Peugeot Citroën and by the European Union within the “SAFEKINEX” Project EVG1-CT-2002-00072. We thank Prof. R. Minetti and Dr. O. Lemaire from the University of Lille for providing us their experimental data.

REFERENCES

(1) Walker, R.W.; Morlay, C., *Oxidation kinetics and autoignition of hydrocarbons,* in *Comprehensive Chemical Kinetics*, vol. 35 (M. Pilling Ed.) Elsevier, 1997.

(2) Klaï, S.E.; Baronnet, F. *J. Chim. Phys.*, **1993**, *90*, 1929-1950.

(3) Klaï, S.E.; Baronnet, F. *J. Chim. Phys.*, **1993**, *90*, 1951-1998.

(4) Voisin, D.; Marchal, A.; Reuillon, M.; Boettner, J.C.; Cathonnet, M. *Combust. Sci. Technol.*, **1998**, *138*, 137-158.

(5) El Bakali, A.; Braun-Unkhoff, M; Dagaut, P.; Frank; P., Cathonnet, M. *Proc. Combust. Inst.*, **2000**, *28*, 1631-1638.

(6) Davis S.G.; Law, C.K. *Combust. Sci. Tech.*, **1998**, *140*, 427-449.

(7) Granata, S.; Faravelli, T.; Ranzi, E. *Combust. Flame*, **2003**, *132*, 533-544.

(8) Lemaire, O.; Ribaucour, M.; Carlier M.; Minetti, R. *Combust. Flame*, **2001**, *127*, 1971-1980.

(9) Buda, F.; Bounaceur, R.; Warth V.; Glaude, P.A.; Fournet, R.; Battin-Leclerc, F. *Combust. Flame*, **2005**, *142*, 170-186.

(10) Touchard, S.; Fournet, R., Glaude, P.A.; Warth, V.; Battin-Leclerc, F.; Vanhove, G.; Ribaucour, M.; Minetti, R. *Proc. Combust. Inst.*, **2005**, *30*, 1073-1081.

(11) Barbé, P., Battin-Leclerc, F.; Côme, G.M. *J. Chim. Phys.* **1995**, *92*, 1666-1692.

(12) Fournet, R.; Bauge, J.C.; Battin-Leclerc, F. *Int. J. Chem. Kin.* **1999**, *31*, 361-379.

(13) Warth, V., Stef, N., Glaude, P.A., Battin-Leclerc, F., Scacchi, G.; Côme, G.M. *Combust. Flame*, **1998**, *114*, 81-102.

(14) Kee, R.J., Rupley, F.M.; Miller, J.A., Sandia Laboratories Report, SAND 89-8009B,1993.

(15) Muller, C.; Michel, V.; Scacchi, G.; Côme, G.M. *J. Chim. Phys.*, **1995**, *92*, 1154-1178.

(16) Benson, S.W., Thermochemical Kinetics, 2nd ed., John Wiley, New York , 1976.

(17) Cohen, N. *J. Phys. Ref. Data*, **1996**, *25*, 1411.

(18) Baulch, D.L.; Cobos, C.J.; Cox, R.A.; Franck, P.; Hayman, G.D.; Just, Th.; Kerr, J.A., Murrells, T.P.; Pilling, M.J.; Troe, J.; Walker, R.W.; Warnatz, J. *Combust. Flame* **1994,** *98***,** 59-79.

(19) Tsang, W.; Hampson, R.F., *J. Phys. Chem. Ref. Data* **1986**, 15, n°3.

(20) Troe , J., *Ber. Buns. Phys. Chem.* **1974**, *78*, 478.

(21) Warth, V.; Battin-Leclerc, F.; Fournet, R.; Glaude, P.A.; Côme, G.M.; Scacchi, G. *Comput. Chem.,* **2000**, *24 (5)* 541-560.

(22) Allara, D.L.; Shaw, R. J. *Phys. Chem. Ref. Data*, **1980**, *9*, 523-559.

(23) Sirjean, B.; Buda, F.; Hakka, H.; Glaude, P.A.; Fournet, R.; Warth, V.; Battin-Leclerc, F.; Ruiz-Lopez, M. *Proc. Combust. Inst.*, **2006**, *31*, in press.

(24) Gulati, S.K; Walker, R.W. J. Chem. Soc. Faraday Trans., **1989**, *85(11)*, 1799-1812.

(25) Pedley, J. B.; Naylor, R.D.; Kirby, S.P., Thermochemical data of organic compounds, $2^{nd}$ ed., Chapman and Hall,1986.

(26) Handford-Styring, S.M.; Walker, R.W. *Phys. Chem. Chem. Phys.*, **2001**, *3*, 2043-2052.

(27) Glaude, P.A.; Battin-Leclerc, F.; Fournet, R.; Warth, V.; Côme, G.M.; Scacchi, G. *Combust. Flame*, **2000**, *122*, 541-462.

(28) Heyberger, B.; Battin-Leclerc, F.; Warth, V.; Fournet, R.; Côme, G.M.; Scacchi, G. *Combust. Flame*, **2001**, *126*, 1780-1802.

(29) Glaude, P.A.; Battin-Leclerc, F.; Judenherc, B.; Warth, V.; Fournet, R.; Côme, G.M.; Scacchi; G.; Dagaut, P.; Cathonnet M. *Combust. Flame*, **2000**, *121*, 345-355.

(30) Frisch, M. J.; Trucks, G. W.; Schlegel, H. B.; Scuseria, G. E.; Robb, M.A.; Cheeseman, J. R.; Montgomery, J. A., Jr.; Vreven, T.; Kudin, K. N.;Burant, J. C.; Millam, J. M.; Iyengar, S. S.; Tomasi, J.; Barone, V.; Mennucci, B.; Cossi, M.; Scalmani, G.; Rega, N.; Petersson, G. A.; Nakatsuji, H.; Hada, M.; Ehara, M.; Toyota, K.; Fukuda, R.; Hasegawa, J.; Ishida, M.; Nakajima, T.; Honda, Y.; Kitao, O.; Nakai, H.; Klene, M.; Li, X.; Knox, J. E.; Hratchian, H. P.; Cross, J. B.; Bakken, V.; Adamo, C.; Jaramillo, J.; Gomperts, R.; Stratmann, R. E.; Yazyev, O.; Austin, A. J.; Cammi, R.; Pomelli, C.; Ochterski, J. W.; Ayala, P. Y.; Morokuma, K.; Voth, G. A.; Salvador, P.; Dannenberg, J. J.; Zakrzewski, V. G.; Dapprich, S.; Daniels, A. D.; Strain, M. C.; Farkas, O.; Malick, D. K.; Rabuck, A.D.; Raghavachari, K.; Foresman, J. B.; Ortiz, J. V.; Cui, Q.; Baboul, A.G.; Clifford, S.; Cioslowski, J.; Stefanov, B. B.; Liu, G.; Liashenko, A.; Piskorz, P.; Komaromi, I.; Martin, R. L.; Fox, D. J.; Keith, T.; Al-Laham, M. A.; Peng, C. Y.; Nanayakkara, A.; Challacombe, M.; Gill, P. M. W.; Johnson, B.; Chen, W.; Wong, M. W.; Gonzalez, C.; Pople, J. A. *Gaussian03*, revision B05; Gaussian, Inc.: Wallingford, CT, 2004.

(31) Dayma, G.; Glaude, P.A.; Fournet, R.; Battin-Leclerc, F. *Int. J. Chem. Kin.*, **2003**, *35*, 273-285.

(32) Da Costa, I.; Fournet, R.; Billaud, F.; Battin-Leclerc, F. *Int. J. Chem. Kin.*, **2003**, *35*, 503-524.

**TABLE 1** : Stucture and thermochemical data of the cyclic species involved in the primary mechanism of the oxidation of cyclohexane as shown in Table 2.

| Species | Structure | $\Delta H_f$ ° (298 K) | S° (298 K) | $C_p$° 300 K | 500 K | 800 K | 1000 K | 1500 K |
|---|---|---|---|---|---|---|---|---|
| $C_6H_{11}$ | | 17.87 | 76.06 | 25.05 | 44.11 | 63.81 | 72.46 | 85.13 |
| $C_6H_{11}OO$ | | -19.44 | 88.73 | 33.64 | 54.20 | 74.88 | 82.87 | 94.57 |
| $C_6H_{10}OOH$-1 | 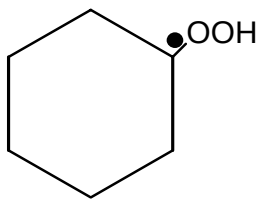 | -11.64 | 94.92 | 35.06 | 54.59 | 74.28 | 82.02 | 93.47 |
| $C_6H_{10}OOH$-2 | 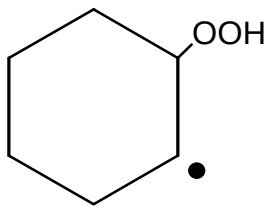 | -7.14 | 95.82 | 35.08 | 54.88 | 74.85 | 82.58 | 93.96 |
| $C_6H_{10}OOH$-3 | 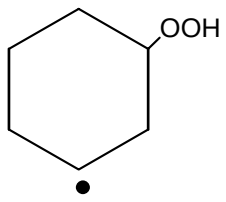 | -7.14 | 95.82 | 35.08 | 54.88 | 74.85 | 82.58 | 93.96 |
| $C_6H_{10}OOH$-4 | 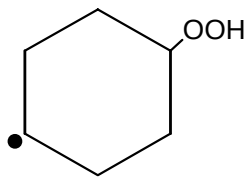 | -7.14 | 94.45 | 35.08 | 54.88 | 74.85 | 82.58 | 93.96 |
| $OOC_6H_{10}OOH$-1 | 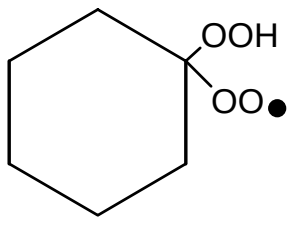 | -47.28 | 101.18 | 39.22 | 62.26 | 84.28 | 92.59 | 105.36 |
| $OOC_6H_{10}OOH$-2 | 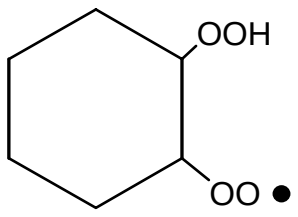 | -43.95 | 108.38 | 41.77 | 63.98 | 84.70 | 92.68 | 104.51 |
| $OOC_6H_{10}OOH$-3 | 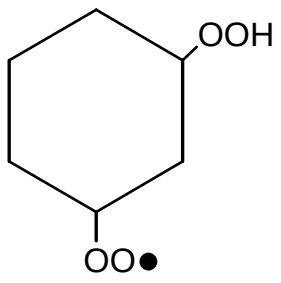 | -43.95 | 108.38 | 41.77 | 63.98 | 84.70 | 92.68 | 104.51 |
| $OOC_6H_{10}OOH$-4 | 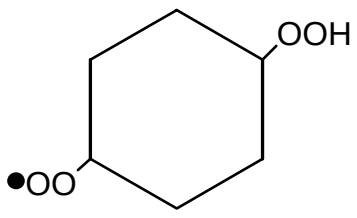 | -43.95 | 108.38 | 41.77 | 63.98 | 84.70 | 92.68 | 104.51 |
| $C_6H_{11}O$ | 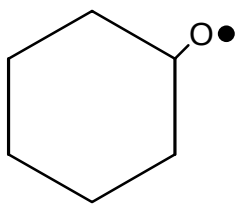 | -17.10 | 75.23 | 28.70 | 48.92 | 69.35 | 78.16 | 90.43 |
| $C_6H_{11}$#6O#3 [a] | 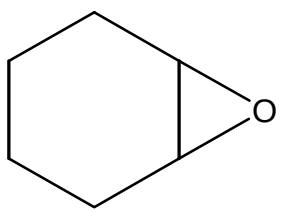 | -28.25 | 74.81 | 30.22 | 50.92 | 70.26 | 77.99 | 91.83 |
| $C_6H_{11}$#6O#5 [b] | 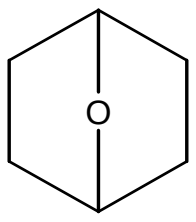 | -42.45 | 74.81 | 30.22 | 50.92 | 70.26 | 77.99 | 91.83 |

[a] Ring strain correction taken equal to that of bicyclo[4,1,0]heptane (17).

[b] Ring strain correction taken equal to that of bicyclo[2,2,1]heptane (17).

**TABLE 2** : Primary mechanism of the oxidation of cyclohexane; only the reactions related to the species containing a 6-members ring are shown.

| Reactions | A[a] | n[a] | Ea[a] | n° |
|---|---|---|---|---|
| **Unimolecular initiations** | | | | |
| H+C6H11<=>C6H12 | $1.0x10^{14}$ | 0.0 | 0.0 | (1) |
| **Bimolecular initiations** | | | | |
| C6H12+O2=>C6H11+HO2 | $8.3x10^{13}$ | 0.0 | 50870 | (2) |
| **Additions to oxygen** | | | | |
| C6H11+O2<=>C6H11OO | $6.0x10^{18}$ | -2.5 | 0.0 | (3) |
| C6H10OOH-1+O2<=>OOC6H10OOH-1 | $6.0x10^{18}$ | -2.5 | 0.0 | (4) |
| C6H10OOH-2+O2<=>OOC6H10OOH-2 | $3.5x10^{18}$ | -2.5 | 0.0 | (5) |
| C6H10OOH-3+O2<=>OOC6H10OOH-3 | $6.0x10^{18}$ | -2.5 | 0.0 | (6) |
| C6H10OOH-4+O2<=>OOC6H10OOH-4 | $6.0x10^{18}$ | -2.5 | 0.0 | (7) |
| **Isomerisations** | | | | |
| C6H11OO<=>C6H10OOH-1 | $9.7x10^{9}$ | 1.0 | 35000 | (8) |
| C6H11OO<=>C6H10OOH-2 | $3.9x10^{10}$ | 1.0 | 35285 | (9) |
| C6H11OO<=>C6H10OOH-3 | $3.9x10^{10}$ | 1.0 | 32800 | (10) |
| C6H11OO<=>C6H10OOH-4 | $9.7x10^{7}$ | 1.0 | 25556 | (11) |
| OOC6H10OOH-1=>OH+C5H9COOOH | $3.9x10^{10}$ | 1.0 | 35285 | (12) |
| OOC6H10OOH-1=>OH+C5H9COOOH | $3.9x10^{10}$ | 1.0 | 32800 | (13) |
| OOC6H10OOH-1=>OH+C5H9COOOH | $9.7x10^{7}$ | 1.0 | 25556 | (14) |
| OOC6H10OOH-2=>OH+C5H9COOOH | $9.7x10^{9}$ | 1.0 | 35000 | (15) |
| OOC6H10OOH-2=>OH+C5H9COOOH | $9.7x10^{9}$ | 1.0 | 30285 | (16) |
| OOC6H10OOH-2=>OH+C5H9COOOH | $3.9x10^{10}$ | 1.0 | 32800 | (17) |
| OOC6H10OOH-2=>OH+C5H9COOOH | $9.7x10^{7}$ | 1.0 | 25556 | (18) |
| OOC6H10OOH-2=>OH+C5H9COOOH | $1.9x10^{10}$ | 1.0 | 35285 | (19) |
| OOC6H10OOH-3=>OH+C5H9COOOH | $9.7x10^{9}$ | 1.0 | 35000 | (20) |
| OOC6H10OOH-3=>OH+C5H9COOOH | $3.9x10^{10}$ | 1.0 | 35285 | (21) |
| OOC6H10OOH-3=>OH+C5H9COOOH | $9.7x10^{9}$ | 1.0 | 27800 | (22) |
| OOC6H10OOH-3=>OH+C5H9COOOH | $9.7x10^{7}$ | 1.0 | 25556 | (23) |
| OOC6H10OOH-3=>OH+C5H9COOOH | $1.9x10^{10}$ | 1.0 | 32800 | (24) |
| OOC6H10OOH-4=>OH+C5H9COOOH | $9.7x10^{9}$ | 1.0 | 35000 | (25) |
| OOC6H10OOH-4=>OH+C5H9COOOH | $3.9x10^{10}$ | 1.0 | 35285 | (28) |
| OOC6H10OOH-4=>OH+C5H9COOOH | $3.9x10^{10}$ | 1.0 | 32800 | (29) |
| OOC6H10OOH-4=>OH+C5H9COOOH | $4.9x10^{7}$ | 1.0 | 20556 | (30) |
| **Beta-scissions** | | | | |
| C6H11<=>CH2=CHCH2CH2CH2CH2 | $4.0x10^{13}$ | 0.0 | 28700 | (31) |
| H+C6H10<=>C6H11 | $2.6x10^{13}$ | 0.0 | 1560 | (32) |
| C6H10OOH-1=>OH+C5H10CO | $1.0x10^{19}$ | 0.0 | 7500 | (33) |
| C6H10OOH-2=>HO2+C6H10 | $8.5x10^{12}$ | 0.0 | 26000 | (34) |
| C6H10OOH-1=>CH2=C(OOH)CH2CH2CH2CH2 | $4.0x10^{13}$ | 0.0 | 28700 | (35) |
| C6H10OOH-2=>CH2=CHCH(OOH)CH2CH2CH2 | $2.0x10^{13}$ | 0.0 | 28700 | (36) |
| C6H10OOH-2=>CH(OOH)=CHCH2CH2CH2CH2 | $2.0x10^{13}$ | 0.0 | 28700 | (37) |
| C6H10OOH-3=>CH2=CHCH2CH(OOH)CH2CH2 | $2.0x10^{13}$ | 0.0 | 28700 | (38) |
| C6H10OOH-3=> CH2=CHCH2CH2CH2CH(OOH) | $2.0x10^{13}$ | 0.0 | 28700 | (39) |
| C6H10OOH-4=>CH2=CHCH2CH2CH(OOH)CH2 | $4.0x10^{13}$ | 0.0 | 28700 | (40) |
| C6H11O=>C5H10CHO | 2.0E+13 | 0.0 | 15000 | (41) |
| **Decompositions in cyclic ethers** | | | | |
| C6H10OOH-2=>OH+C6H10#6O#3[b] | 2.06E+13 | 0.0 | 9680 | (42) |
| C6H10OOH-3=>OH+C5H9CHO | 2.06E+13 | 0.0 | 19590 | (43) |
| C6H10OOH-4=>OH+C6H10#6O#5[b] | 2.06E+13 | 0.0 | 17100 | (44) |
| **Oxidations** | | | | |
| C6H11+O2=>C6H10+HO2 | $3.9x10^{12}$ | 0.0 | 5000 | (45) |
| **Metatheses** | | | | |
| C6H12+H=>C6H11+H2 | $2.68x10^{8}$ | 2.0 | 5000 | (46) |
| C6H12+OH=>C6H11+H2O | $0.77x10^{8}$ | 2.0 | -770 | (47) |
| C6H12+HO2=>C6H11+H2O2 | $1.20x10^{13}$ | 0.0 | 15500 | (48) |
| C6H12+CH3=>C6H11+CH4 | $1.20x10^{12}$ | 0.0 | 9600 | (49) |
| C6H12+CHO=>C6H11+HCHO | $6.44x10^{7}$ | 1.9 | 17000 | (50) |
| C6H12+CH2OH=>C6H11+CH3OH | $3.62x10^{2}$ | 2.95 | 12000 | (51) |
| C6H12+CH3O=>C6H11+CH3OH | $8.69x10^{11}$ | 0.0 | 4500 | (52) |
| C6H12+CH3OO=>C6H11+CH3OOH | $1.80x10^{13}$ | 0.0 | 17500 | (53) |
| C6H12+C2H5=>C6H11+C2H6 | $1.20x10^{12}$ | 0.0 | 11000 | (54) |
| C6H12+iC3H7=>C6H11+C3H8 | $1.20x10^{12}$ | 0.0 | 12200 | (55) |
| C6H12+C6H11OO=>C6H11+C6H11OOH | $1.80x10^{13}$ | 0.0 | 17500 | (56) |

| **Disproportionations** | | | | |
|---|---|---|---|---|
| C6H11OO+HO2=>C6H11OOH+O2 | $2.0x10^{11}$ | 0.0 | -1300 | (57) |
| C6H11OO+C6H11OO=>C5H10CO+C6H11OH+O2 | $1.4x10^{10}$ | 0.0 | -725 | (58) |
| 2C6H11OO=>2C6H11O+O2 | $6.30x10^{10}$ | 0.0 | -725 | (59) |
| C6H11OO+CH3OO=>C5H10CO+CH3OH+O2 | $1.4Ex10^{10}$ | 0.0 | -725 | (60) |
| C6H11OO+CH3OO=>C6H11OH+HCHO+O2 | $1.4Ex10^{10}$ | 0.0 | -725 | (61) |
| C6H11OO+C2H5OO=>C5H10CO+C2H5OH+O2 | $1.4Ex10^{10}$ | 0.0 | -725 | (62) |
| C6H11OO+C2H5OO=>C6H11OH+CH3CHO+O2 | $1.4Ex10^{10}$ | 0.0 | -725 | (63) |
| C6H11OO+iC3H7OO=>C5H10CO+C3H7OH+O2 | $1.4Ex10^{10}$ | 0.0 | -725 | (64) |
| C6H11OO+iC3H7OO=>C6H11OH+C2H5CHO+O2 | $1.4Ex10^{10}$ | 0.0 | -725 | (65) |

[a] The rate constants are given at 1 atm ($k=A\ T^n \exp(-E_a/RT)$) in $cm^3$, mol, s, cal units.

[b] A species finishing by #x is a cyclic ethers containing both a 6-members ring and x-members ether ring.

**TABLE 3** : Rate parameters for the isomerizations of peroxycyclohexyl radicals.

| Size of the cycle of the transition state | Model Transition state | Strain energy ($SE_{ts}$) of the transition state (kcal/mol) | A ($s^{-1}$) | Ea (kcal/mol) | Ea (kcal/mol) for acyclic compounds | $k_{753K}$ ($s^{-1}$) | $k_{753K}$ ($s^{-1}$) (26) |
|---|---|---|---|---|---|---|---|
| 6 | 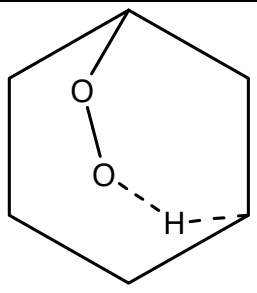 | 15.8 | $3.9 \times 10^{10} \times T$ | 32.8 | 25.0 | $6.6 \times 10^{3}$ | $7.1 \times 10^{3}$ |
| 4 | 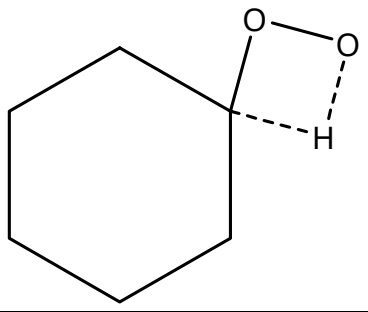 | 23 | $9.7 \times 10^{9} \times T$ | 35.0 | 35.0 | $5.1 \times 10^{2}$ | - |
| 5 | 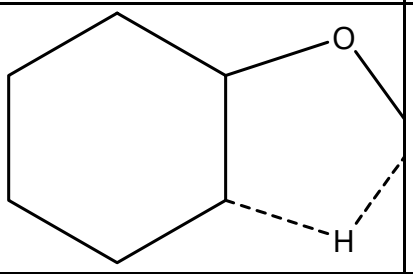 | 18.3 | $3.9 \times 10^{10} \times T$ | 35.3 | 32.5 | $1.7 \times 10^{3}$ | $1.3 \times 10^{3}$ |
| 7 | 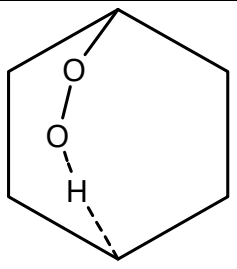 | 7.5 | $9.7 \times 10^{7} \times T$ | 25.5 | 22.0 | $2.9 \times 10^{3}$ | $2.3 \times 10^{3}$ |

**TABLE 4** : Rate parameters for the formation of cyclic ethers.

| Size of the cycle of the transition state | A ($s^{-1}$) | Ea (kcal/mol) |
|---|---|---|
| 3 | $2.06x10^{13}$ | 9.68 |
| 4 | $2.06x10^{13}$ | 19.6 |
| 5 | $2.06x10^{13}$ | 17.1 |

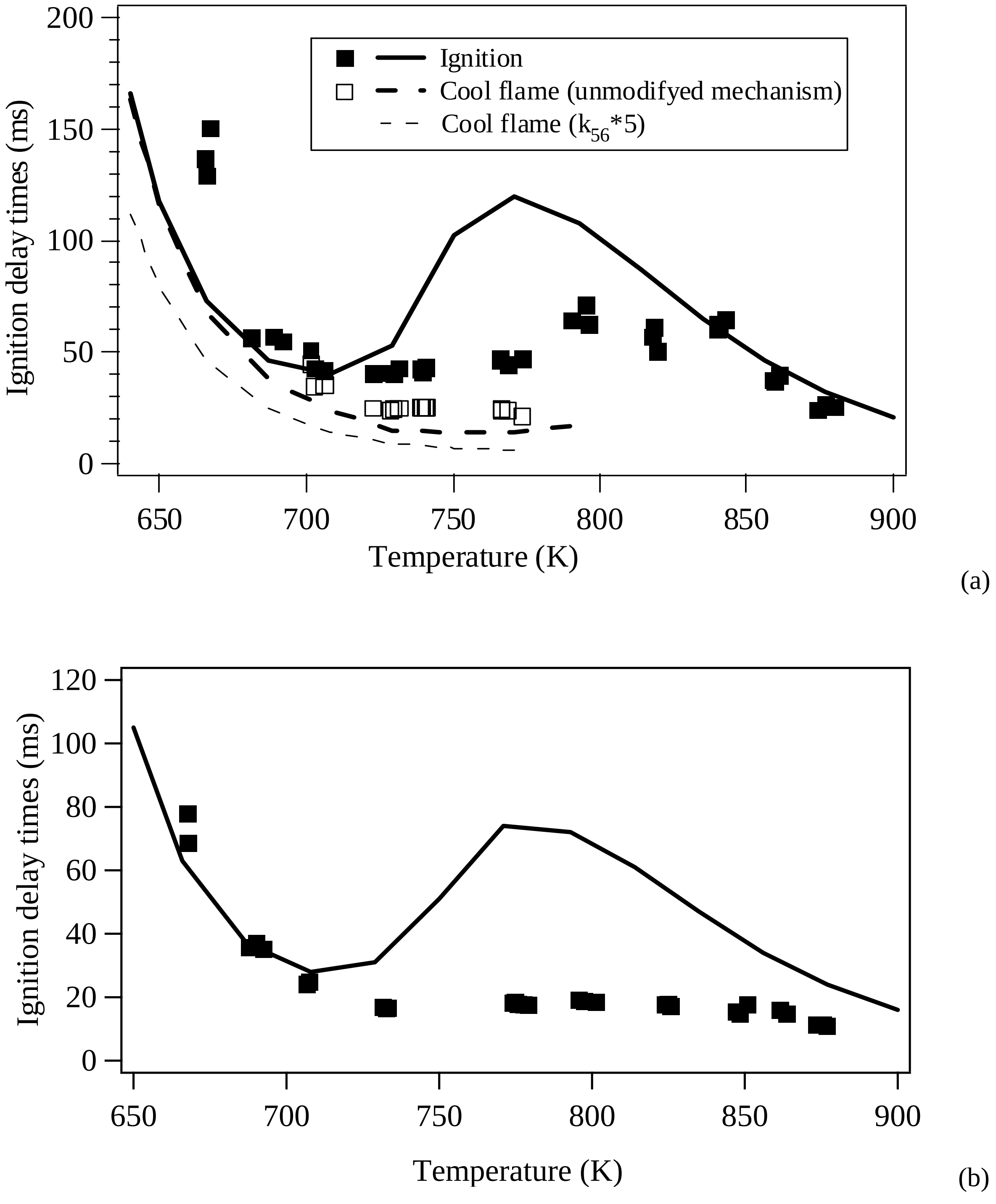

Figure 1: Auto-ignition and cool flame delay times vs. temperature in a rapid compression machine for an equivalence ratio of 1 at a pressure after compression (a) from 7.2 to 9.2 bar (initial pressure of 0.467 bar) and (b) from 11.4 to 14.3 bar (initial pressure of 0.730 bar) (8).

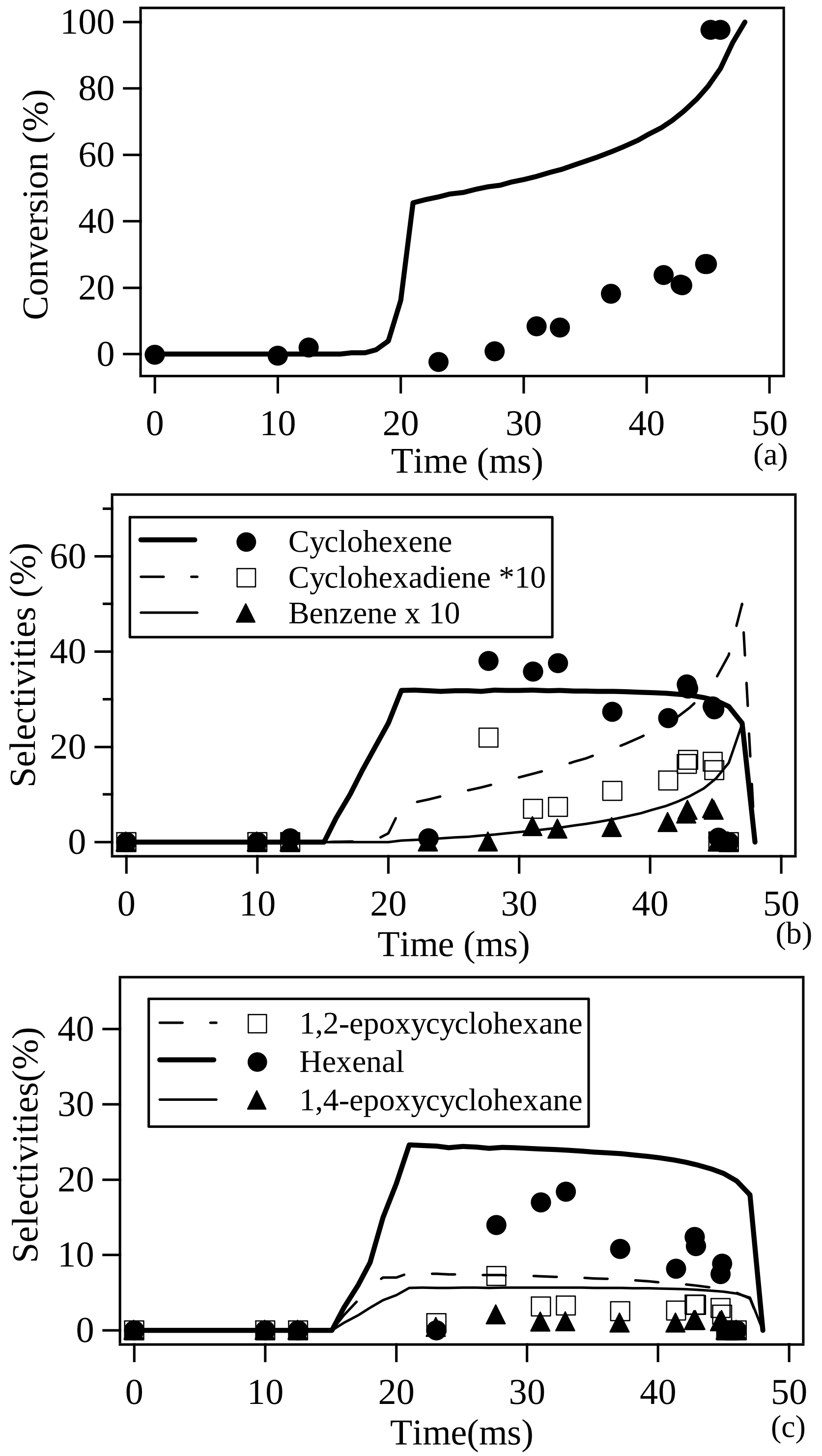

Figure 2 : Conversion of reactant (a) and selectivities (b, c) of main products vs. reaction time during the oxidation of cyclohexane a in rapid compression machine under the conditions of figure 1a at 722 K (8).

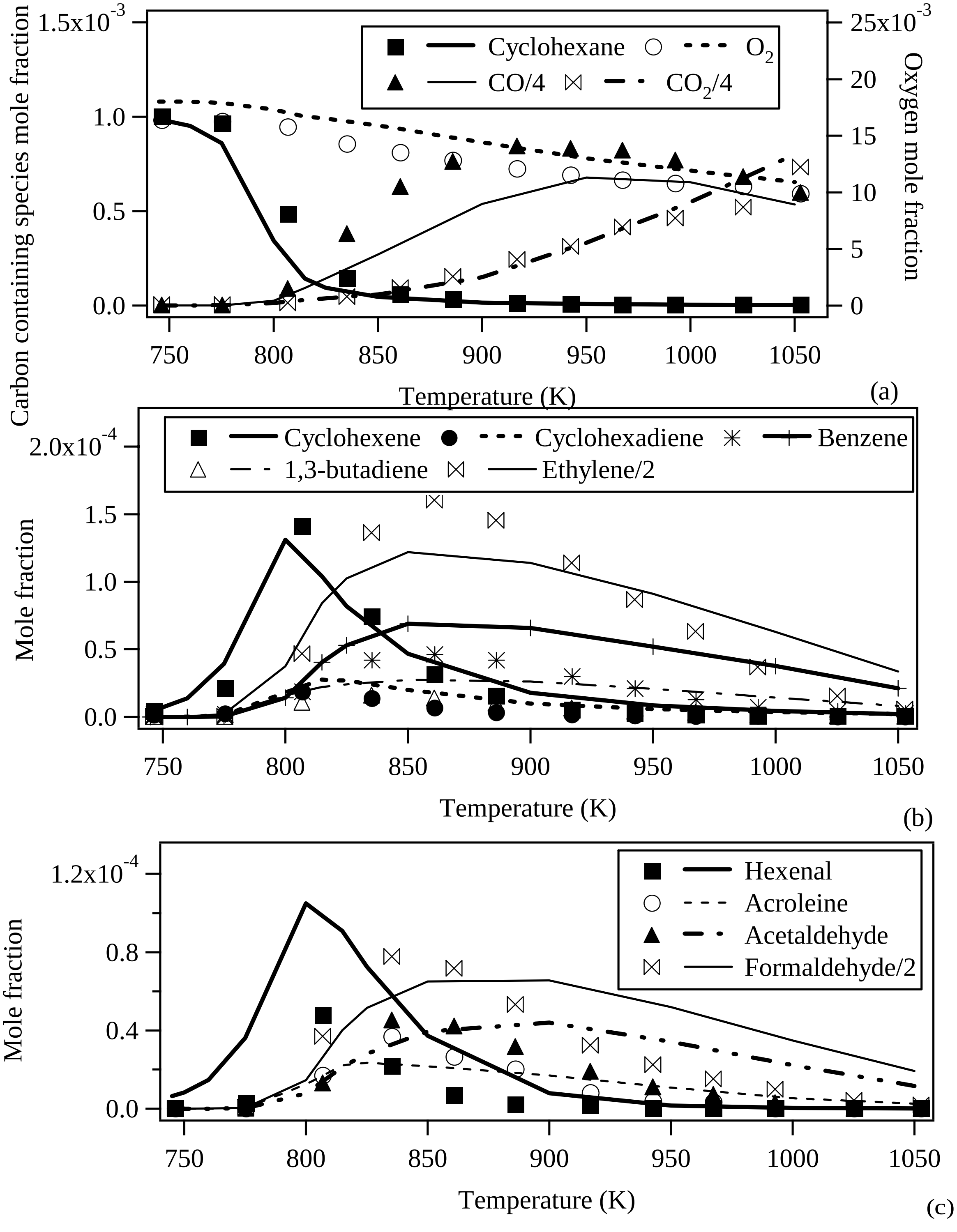

Figure 3: Profiles of products in a jet-stirred reactor at 10 bar, $\phi$ = 0.5 and a residence time of 0.5 s (4).

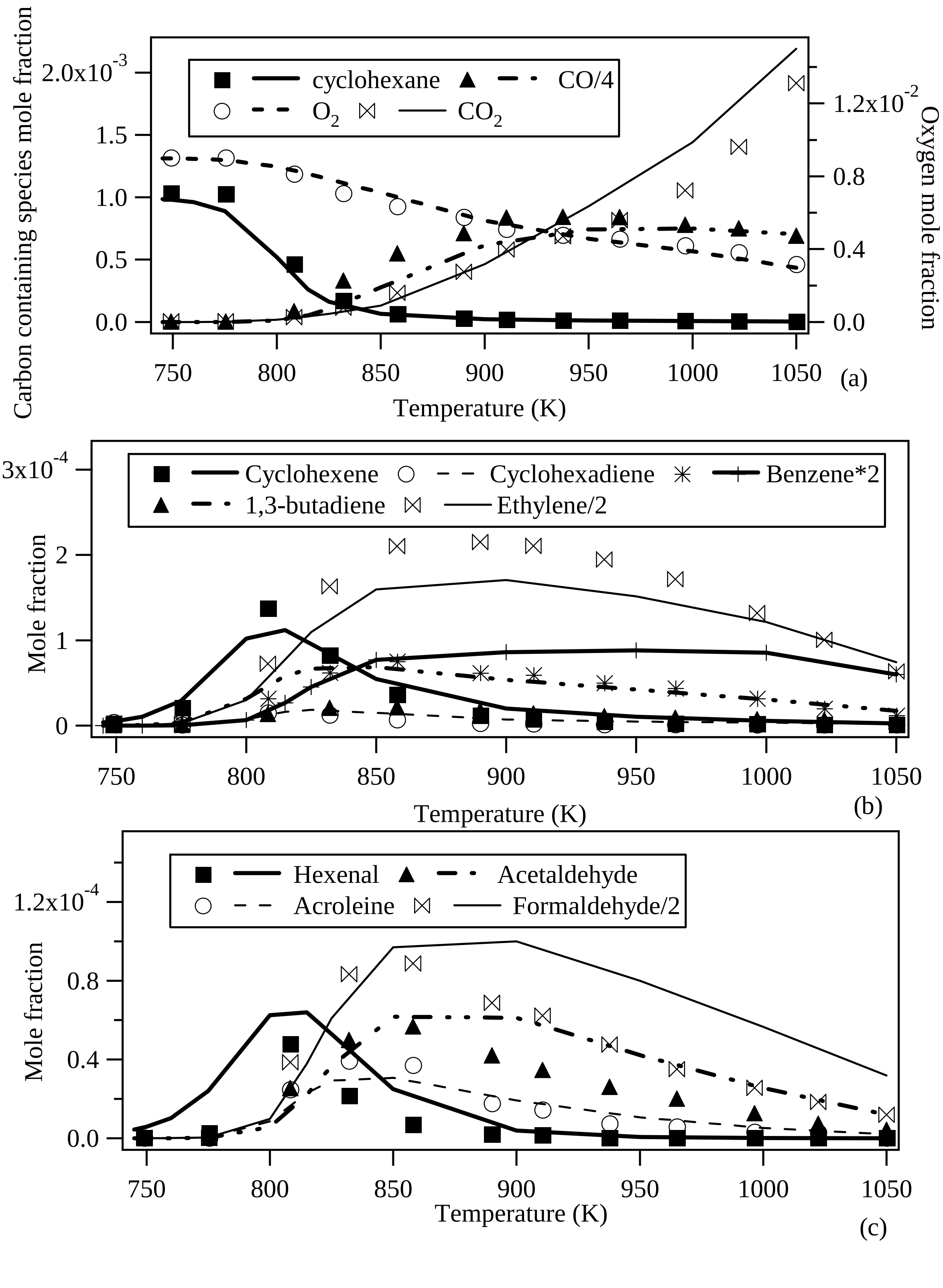

Figure 4: Profiles of products in a jet-stirred reactor at 10 bar, $\phi$ = 1 and a residence time of 0.5 s (4).

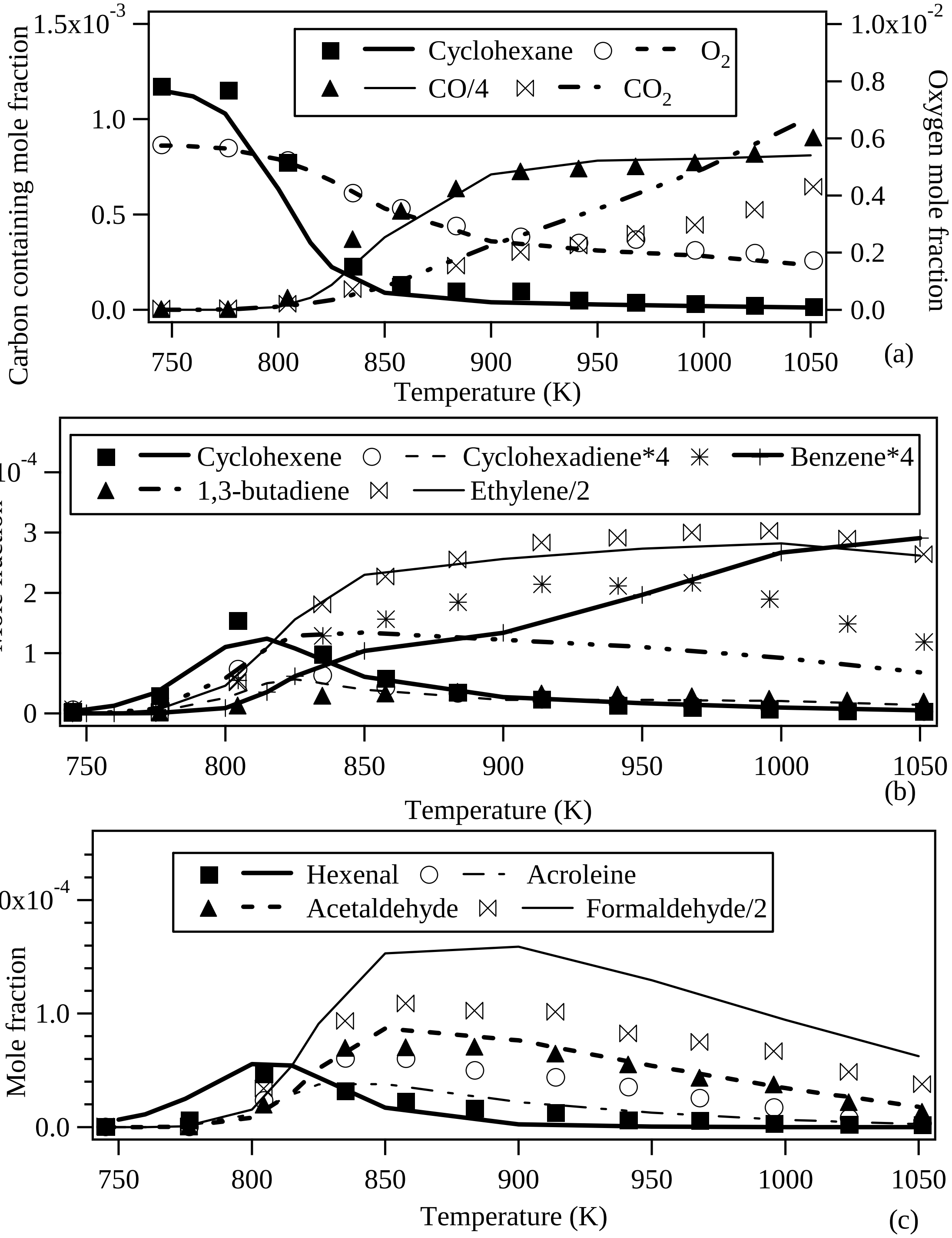

Figure 5: Profiles of products in a jet-stirred reactor at 10 bar, $\phi$ = 1.5 and a residence time of 0.5 s (4).

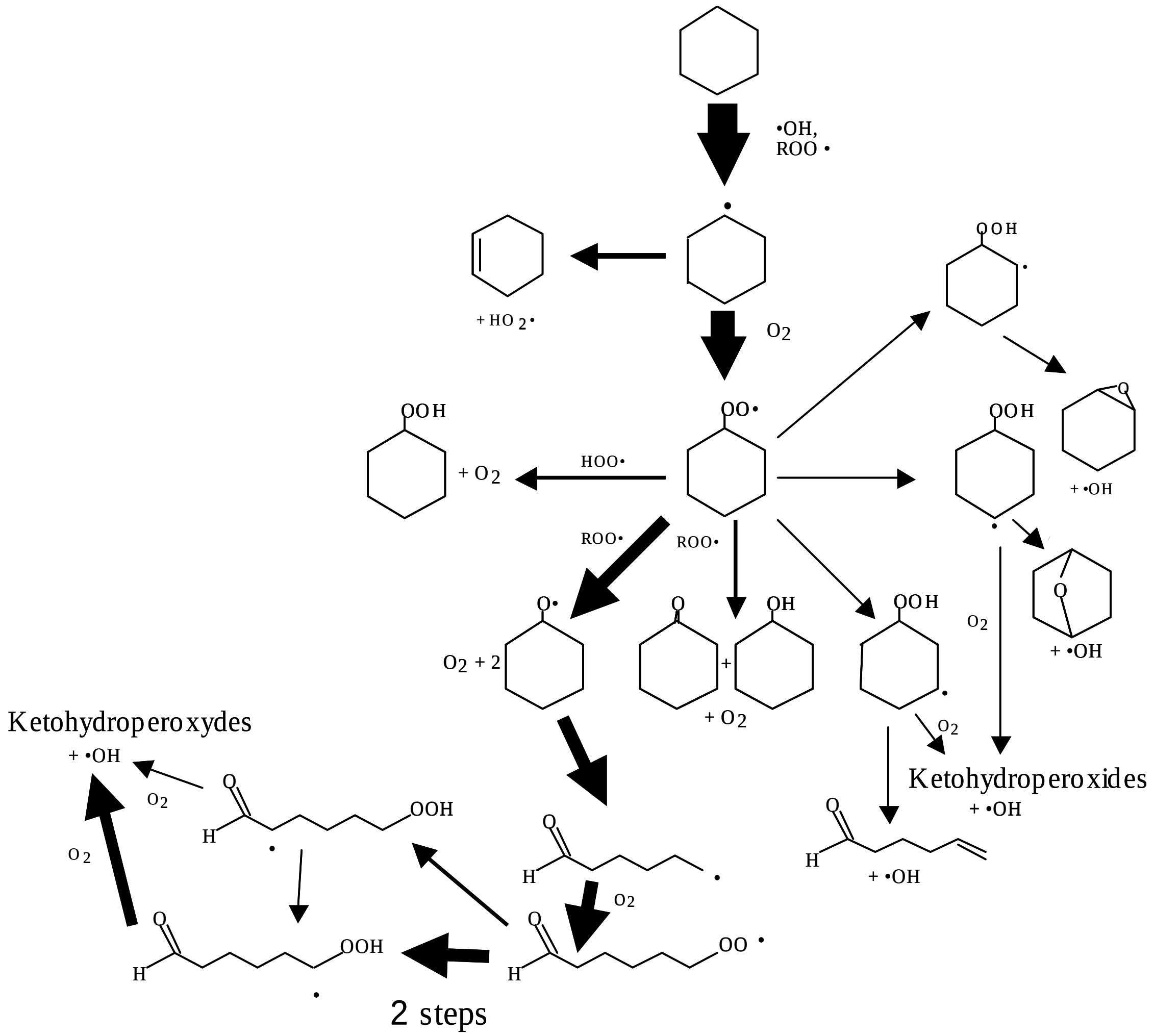

Figure 6: Flow rates analyses under the conditions of figure 1a at 650 K at 20 % conversion. The thickness of the arrows is proportional to the flow rate of the related reaction.

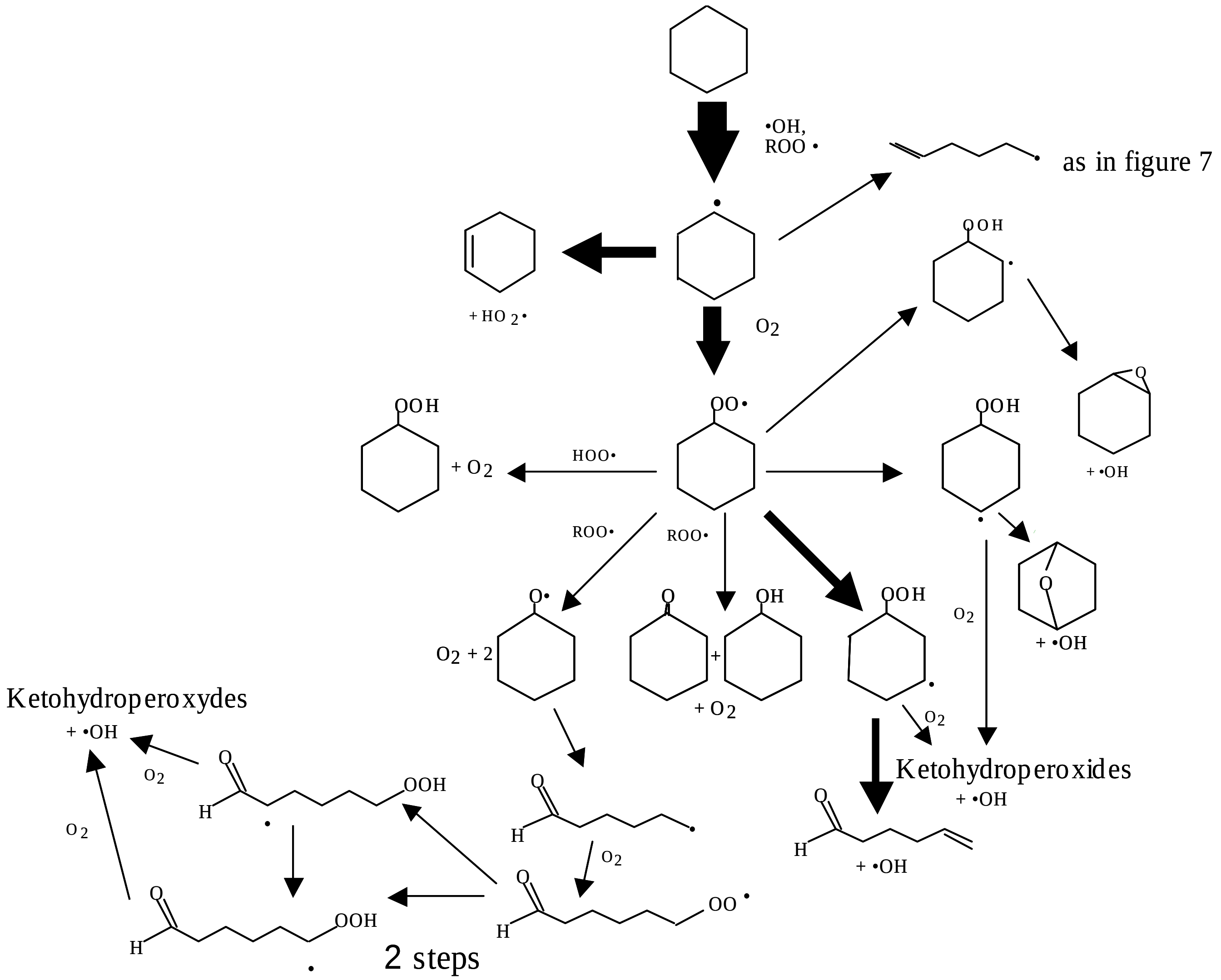

Figure 7 : Flow rates analyses under the conditions of figure 1a at 750 K at 20 % conversion. The thickness of the arrows is proportional to the flow rate of the related reaction.

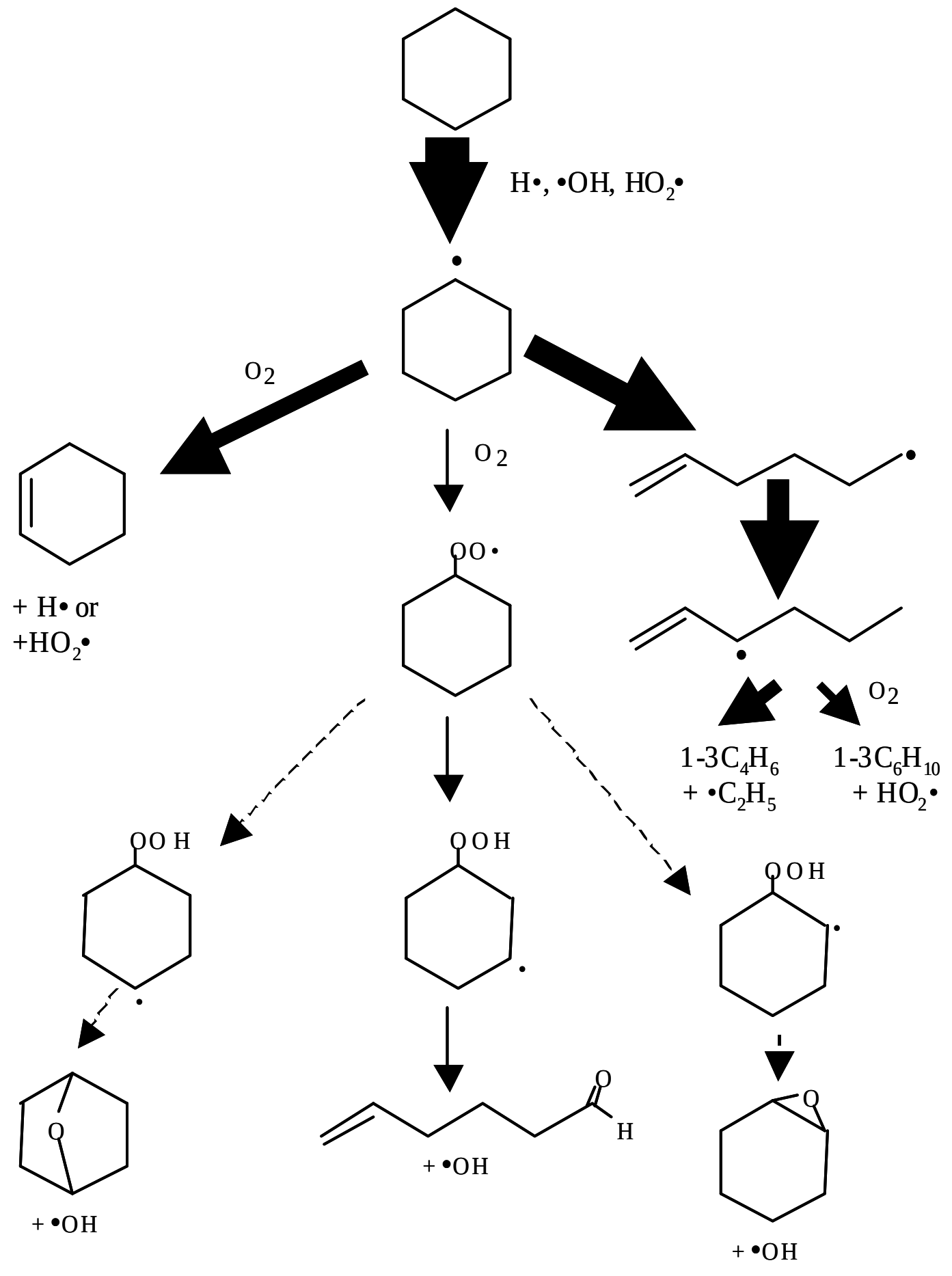

Figure 8: Flow rates analysis under the conditions of figure 4 at 900 K for 20 % conversion corresponding to a residence time of 0.03 s. The thickness of the arrows is proportional to the flow rate of the related reaction.

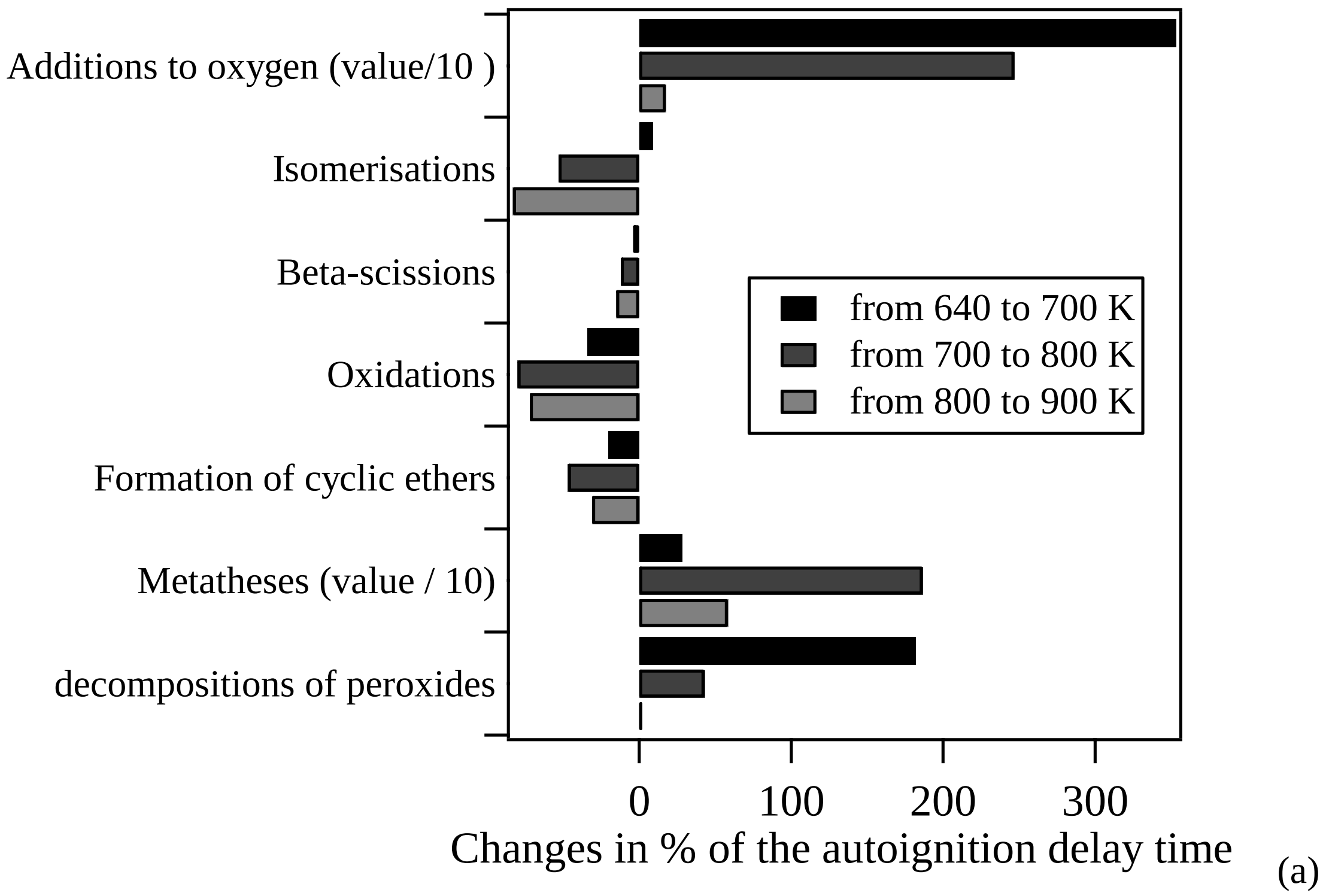

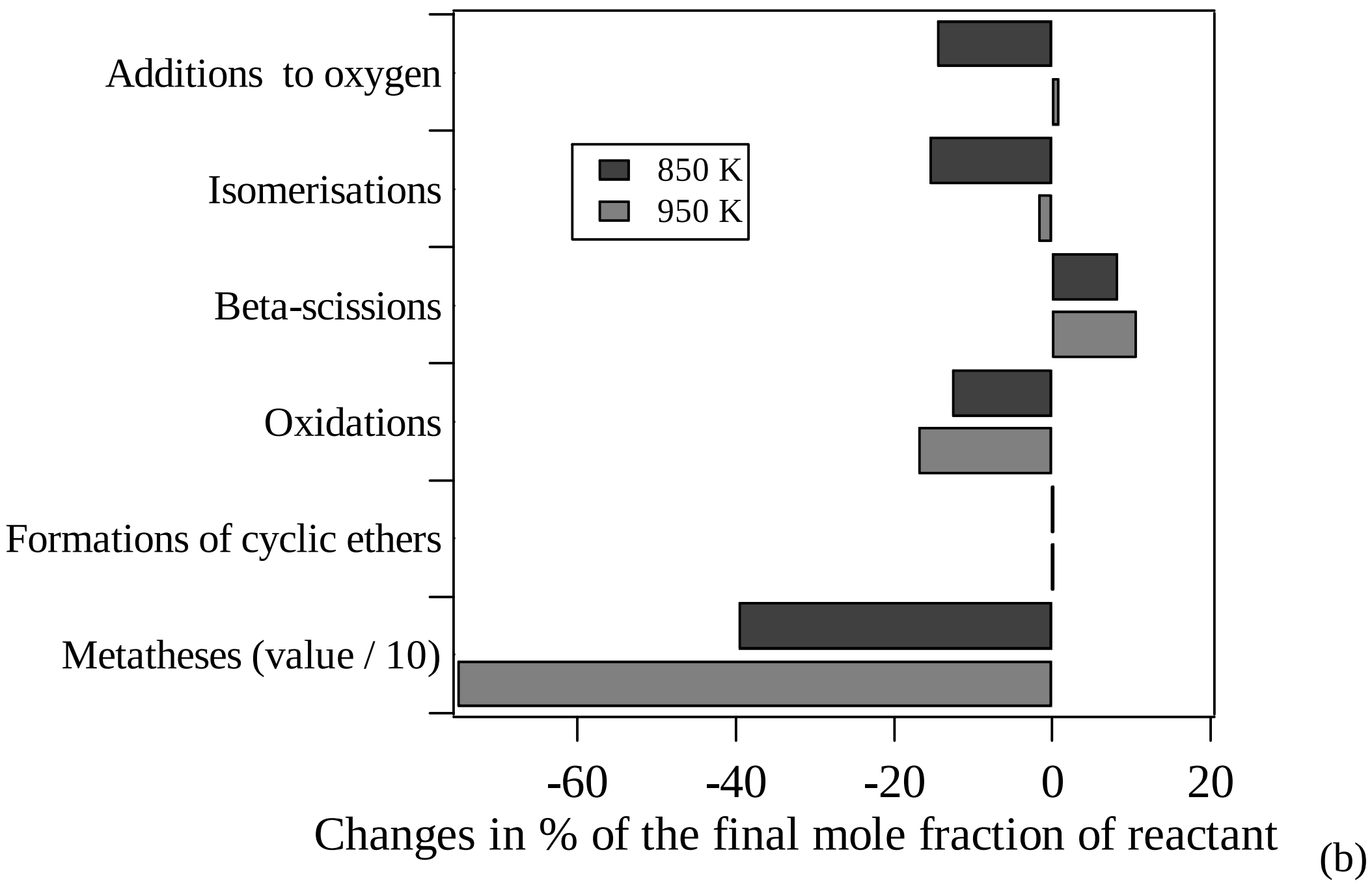

Figure 9: Sensitivity analysis computed under the conditions of (a) figure 1a and (b) figure 4: the rate constant of each presented generic reaction has been divided by a factor 10. Only reactions for which a change above 5 % has been obtained are presented. To fit in the figure, the changes in % obtained for some very sensitive reactions have been divided by a factor 10.